\documentclass[twocolumn,floatfix,noeprint]{revtex4-2}
\usepackage[utf8]{inputenc}
\usepackage[caption=false]{subfig}
\usepackage{graphicx,tabularx,makecell}
\usepackage{amsmath,amsfonts,amssymb}
\usepackage{mathtools}
\usepackage{hhline}
\usepackage{bm,dsfont}
\usepackage[dvipsnames]{xcolor}
\usepackage[colorlinks=true, citecolor=Green, linkcolor=BrickRed, urlcolor=NavyBlue]{hyperref}
\usepackage[all]{hypcap}
\usepackage[T5,T1]{fontenc}
\usepackage[normalem]{ulem}
\usepackage[english]{babel}

\usepackage{ulem}
\usepackage{xcolor}

\newcommand{\editor}[2]{%
  \expandafter\newcommand\csname #1note\endcsname[1]{%
    \textcolor{#2}{(\textbf{#1:} \textit{##1})}}%
  \expandafter\newcommand\csname #1\endcsname[1]{%
    \textcolor{#2}{##1}}%
  \expandafter\newcommand\csname #1cancel\endcsname[1]{%
    \textcolor{#2}{\sout{##1}}}%
  \expandafter\newcommand\csname #1change\endcsname[2]{%
    \textcolor{#2}{\sout{##1} ##2}}%
  \newenvironment{#1text}{\color{#2}}{\color{black}}
}

\editor{CS}{blue}

\graphicspath{{}}

\begin{document}

\title{Machine-Learning Interatomic Potential for Twisted Hexagonal Boron Nitride: Accurate Structural Relaxation and Emergent Polarization}

\author{Wilson Nieto Luna}
\email{wilson.nieto@uantwerpen.be}
\affiliation{Department of Physics and NANOlight Center of Excellence, University of Antwerp, Groenenborgerlaan 171, 2020 Antwerp, Belgium}
\author{Robin Smeyers}
\affiliation{Department of Physics and NANOlight Center of Excellence, University of Antwerp, Groenenborgerlaan 171, 2020 Antwerp, Belgium}
\author{Lucian Covaci}
\affiliation{Department of Physics and NANOlight Center of Excellence, University of Antwerp, Groenenborgerlaan 171, 2020 Antwerp, Belgium}
\author{Cem Sevik}
\affiliation{Department of Physics and NANOlight Center of Excellence, University of Antwerp, Groenenborgerlaan 171, 2020 Antwerp, Belgium}
\author{Milorad V. Milo\v{s}evi\'c}
\affiliation{Department of Physics and NANOlight Center of Excellence, University of Antwerp, Groenenborgerlaan 171, 2020 Antwerp, Belgium}

\date{\today}

\begin{abstract}

The emerging ferroelectric properties of two-dimensional (2D) heterostructures are at the forefront of science and prospective technology. In moir\'{e} bilayers, twisting or heterostructuring causes local atomic reconstruction, which even at picometer scale, can lead to pronounced ferroelectric polarization. Accurately determining this reconstruction utilizing ab initio methods is unfeasible for the relevant system sizes, but modern machine-learning interatomic potentials offer a viable solution. Here, we present the Gaussian Approximation Potential for twisted hexagonal boron nitride (hBN) layers validated against ab initio datasets. This approach enables the precise analysis of their structural properties, which is particularly relevant at small twist angles. We couple the structural information to a tight-binding model based on accurate interatomic positioning, and determine the twist-dependent polarization, yielding results that closely align with previous experimental findings - even at room temperature. This methodology enables further studies that are unattainable otherwise and is transferable to other 2D materials of interest.

\end{abstract}

\maketitle

Ferroelectric materials exhibit spontaneous electric polarization that can be controlled by an external electric field. This polarization originates from a symmetry-breaking structural phase transition, leading to a spontaneous alignment of electric dipoles that gives rise to a macroscopic polarization \cite{liu_van_2019, vizner_stern_interfacial_2021, woods_charge-polarized_2021}. Research on controlling ferroelectricity has advanced rapidly in recent years due to its relevance in diverse fields including electronics \cite{mikolajick_next_2021}, telecommunications \cite{mears_telecommunications_1996}, and energy-related applications \cite{sharma_lead-free_2021}. Traditional approaches, such as thinning bulk oxide ferroelectrics \cite{higashitarumizu_purely_2020}, have proven to be challenging. However, the emergence of two-dimensional (2D) materials and their van der Waals (vdW) heterostructures offers a feasible alternative for designing ferroelectrics with added versatility. Specifically, properties that are highly sensitive to strain \cite{brehm_tunable_2020, wu_twisted-layer_2024}, bending \cite{island_gate_2015}, stacking order \cite{yuan_room-temperature_2019}, and interlayer twisting \cite{zhang_twisted_2023} are highly relevant for ferroelectric properties. Furthermore, vdW materials can be integrated with existing metal-oxide semiconductor technology \cite{mehta_depolarization_1973}, making them promising candidates for post-Moore’s law nanoelectronics.  

Intrinsic ferroelectricity in vdW materials is scarce because it requires the material’s bulk crystal to belong to a polar space group \cite{zhang_ferroelectric_2023}. Furthermore, multilayer vdW crystals, such as graphene or hexagonal boron nitride ($h$-BN), tend to form stable centrosymmetric stacking configurations, naturally excluding ferroelectricity. To overcome this drawback, potential vdW ferroelectrics have been engineered by introducing a twist between layers, thereby breaking the central symmetry of the heterostructure \cite{yasuda_stacking-engineered_2021}. Recent experiments have validated this concept by revealing highly polarized domains in twisted $h$-BN\cite{yasuda_stacking-engineered_2021}, revealing  a crucial building block for high-density, non-volatile memory applications \cite{khan_future_2020}. Subsequent experiments have expanded this approach, driven not only by the potential of twisted 2D materials to act as novel polarized materials but also by the unique functionality provided by periodic polarization patterns. These patterns create tunable spatially varying  electrostatic potentials, which can dynamically modulate the electronic\cite{ding_engineering_2024}, optical\cite{roux_optical_2025}, and magnetic properties of adjacent 2D materials. Such capabilities open new avenues for engineering quantum phenomena -- including charge density wave manipulation\cite{tilak_proximity_2024} or exciton trapping\cite{cho_moire_2025, aggoune_dimensionality_2018, zhao_universal_2021} --  in tailored two-dimensional heterostructures.

A comprehensive computational investigation of these fundamentally and technologically promising features is essential; however, it is complicated due to the large-scale periodicity inherent in twisted structures, which prevents atomic relaxation at the first-principles level. Whereas, precise polarization characterization requires highly accurate alignment of the out-of-plane distances among atoms constituting layers. Empirical potentials offer a partial solution for computational cost, but the accuracy level for precise polarization prediction is highly questionable.  However, a state-of-the-art class of potentials based on machine learning (ML) methodologies has recently emerged, advancing the accuracy of energy and force up to the first-principles levels. 

Kernel-based ML potentials\cite{deringer_machine_2019} describe the potential energy surface (PES) of a system as a sum of local environments, ensuring invariance to translation, rotation, and permutation of atoms\cite{zuo_performance_2020}. In this framework, Gaussian Approximation Potentials (GAP),  \cite{bartok_representing_2013, deringer_gaussian_2021}, apply Gaussian process regression to interpolate the PES, making them a robust tool for accurately predicting atomic energies and forces. This methodology has demonstrated success in modeling the properties of materials such as silicon, gold, and carbon\cite{rowe_accurate_2020}, and has shown promising results in systems like graphene\cite{rowe_development_2018} and hexagonal boron nitride (hBN)\cite{thiemann_machine_2020, kocabas_gaussian_2023}.

Considering these aspects, to identify the polarization in twisted $h$-BN heterobilayer crystals, we first developed a GAP-based machine learning model to capture interatomic forces with first-principles accuracy, addressing both in-plane and out-of-plane interactions. Subsequently, using a tight-binding (TB) approach, we calculated the intrinsic polarization of twisted $h$-BN crystals. The results demonstrate that the complete relaxation of ions in twisted layers significantly influences the final polarization values, and that an accurate representation of atomic positions in these layers yields excellent agreement with experimental predictions.

\begin{figure}[!h]
    \centering
    \includegraphics[width=\linewidth]{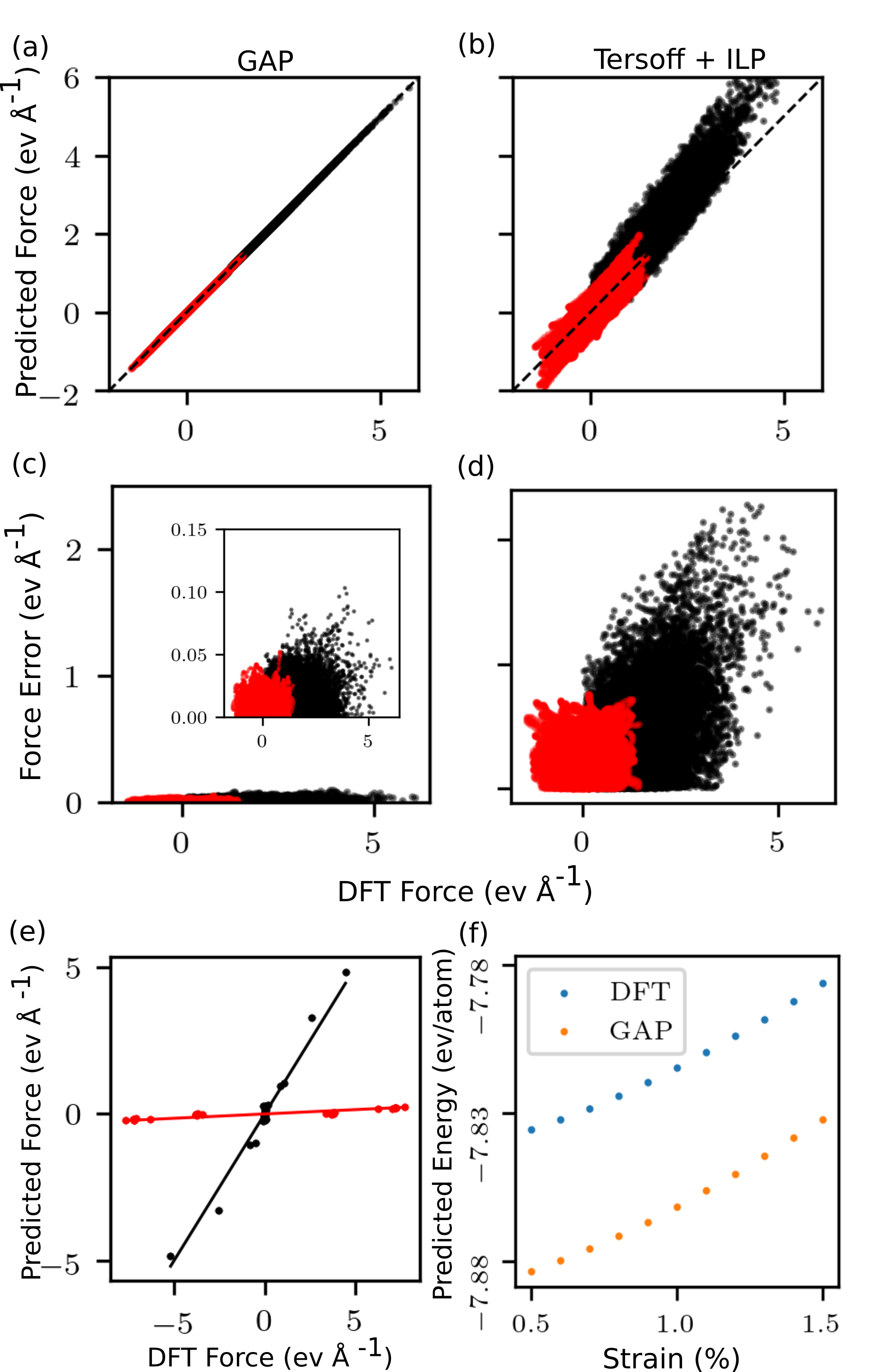}
    \caption{ (a)Force correlation plot for twisted $h-BN$ using the GAP potential. (b) Force correlation plot for twisted $h-BN$ using the classical potentials. (c) Error distribution  for the GAP potential, the inset shows a zoom in view of the distribution of the errors.(d) Error distribution for the classical potential. (e) Force correlation plot for a twisted system of $h$-BN with 1\% uniaxial strain. (f) Comparison of predicted energy between ab-initio methods and our model for different values of strain. In all the pictures, red dots indicate out-of-plane forces and black dots indicate in-plane forces. }
    \label{fig:fig1}
\end{figure}

We start by constructing the ML potential for twisted $h$-BN in order to tackle the problem of atomic relaxation in this type of structures. To construct an accurate database, we followed a similar procedure as the one described in \cite{rowe_development_2018}, in which we build the training configuration with an iterative approach. Training configurations were obtained from first-principles calculations on twisted structures with angles $\theta$ between 21.75$^\circ$ and 7.34$^\circ$, based on DFT as implemented by the Vienna Ab initio Simulation Package (VASP) \cite{kresse_ab_1994, kresse_efficient_1996}. With a reliable database, we can build an accurate interatomic potential based on the DFT energies and forces. From this potential we can predict structural properties of smaller twist angles that were not included in the training configurations. A more detailed explanation about the training configuration is presented in the Supporting Information. The main indication of the quality of the potential is the quality of the forces it predicts compared to a proper reference. In our case we will compare forces predicted by our model in the validation set to the original values obtained from a DFT calculation. This is shown in fig.\ref{fig:fig1}.a, we can see the correlation between forces predicted by the GAP model and the DFT calculation align very closely, with the errors being presented in fig.\ref{fig:fig1}.c. We divide the forces into the out-of-plane and in-plane component, shown in red and black respectively in fig.\ref{fig:fig1}.a-d. The forces were obtained with an RMSE of 0.012 eV \r{A}$^{-1}$ in the out-of-plane direction, and 0.019 eV \r{A}$^{-1}$ in the in-plane direction, where the errors follow a Gaussian distribution as expected. In figure \ref{fig:fig1}.b we show the same force correlation and distribution of errors for the Tersoff\cite{sevik_characterization_2011, sevik_influence_2012, kinaci_thermal_2012} potential in combination with the ILP potential \cite{ouyang_nanoserpents_2018}, the current state of the art classical potential used for this type of system. Our model shows a significant improvement in the prediction of the interlayer interaction, offering an improvement of about one order of magnitude in the prediction of forces. In fig.\ref{fig:fig1}.e-f we present the correlation between the predicted forces of our GAP potential and the DFT forces for a bilayer system with 1\% uniaxial strain in the $y$-axis, and we also compare the energy predicted by DFT and GAP methods. Although these types of systems were not included in the training configuration, the twisted structures present intrinsic strain that was captured by the GAP potential and it is able to predict the forces and energy, as shown in the correlation plot, with an RMSE of 0.24 eV \r{A}$^{-1}$ and 0.011 eV/atom respectively. This accuracy can be improved by including strained structures in our training set.

\begin{figure*}[t]
    \centering
    \includegraphics[width=\linewidth]{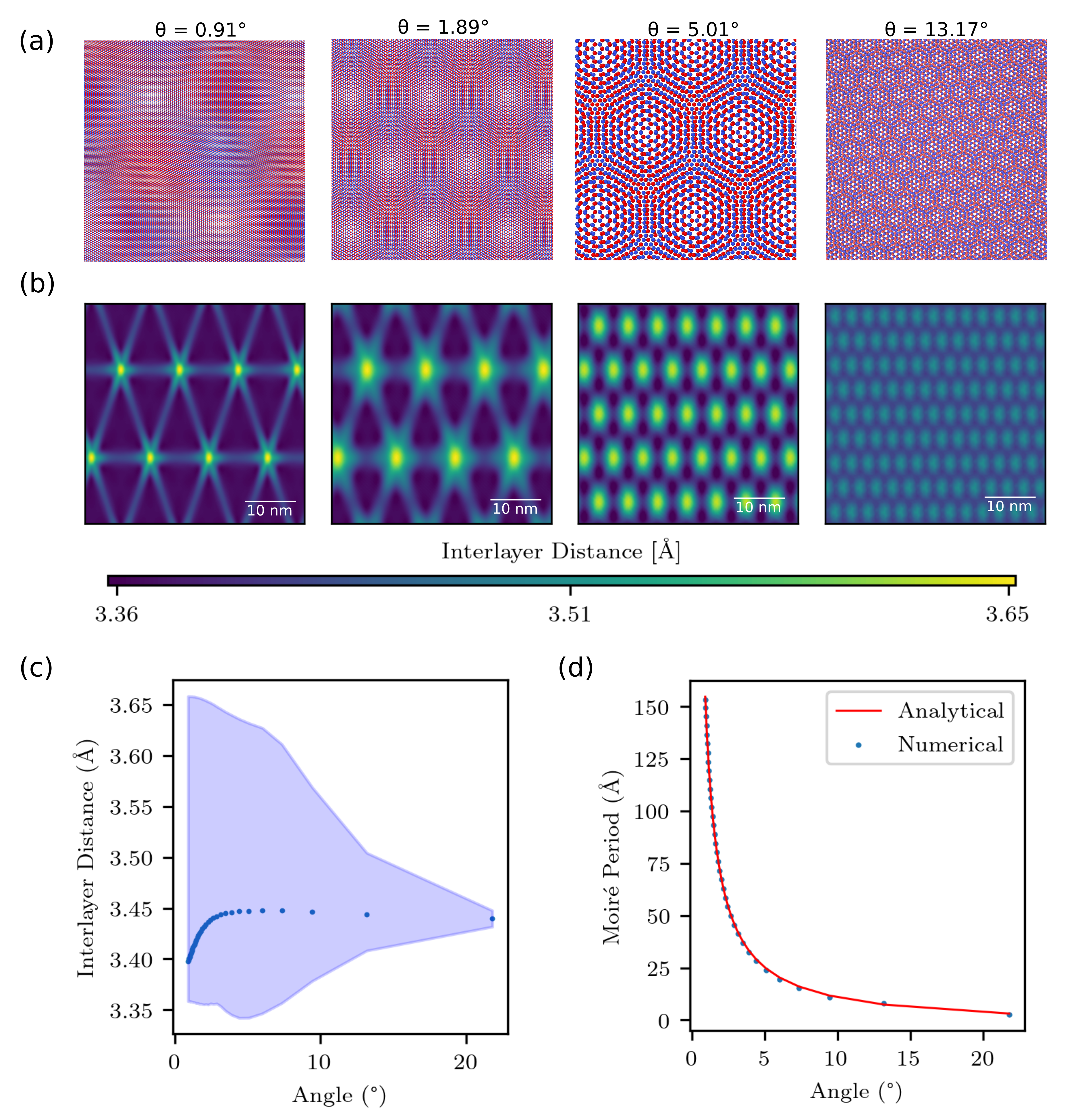}
    \caption{Different structures with corresponding $\theta$ for different regimes in twisted bilayer hBN. (a) Top view of the relaxed lattice with B atoms in red and N atoms in blue. (b) Interlayer distance landscape, calculated from the atoms in the top layer. (c) $d_{max}$, $d_{min}$ and $\langle d \rangle$ as a function of the twist angle. The blue dots represent $\langle d \rangle$ and the blue shade represents the interval between $d_{max}$ and $d_{min}$. (d) Moir\'{e} period as a function of the twist angle.}
    \label{fig:interlayer_distance}
\end{figure*}

\begin{figure*}[t]
    \centering
    \includegraphics[width=\linewidth]{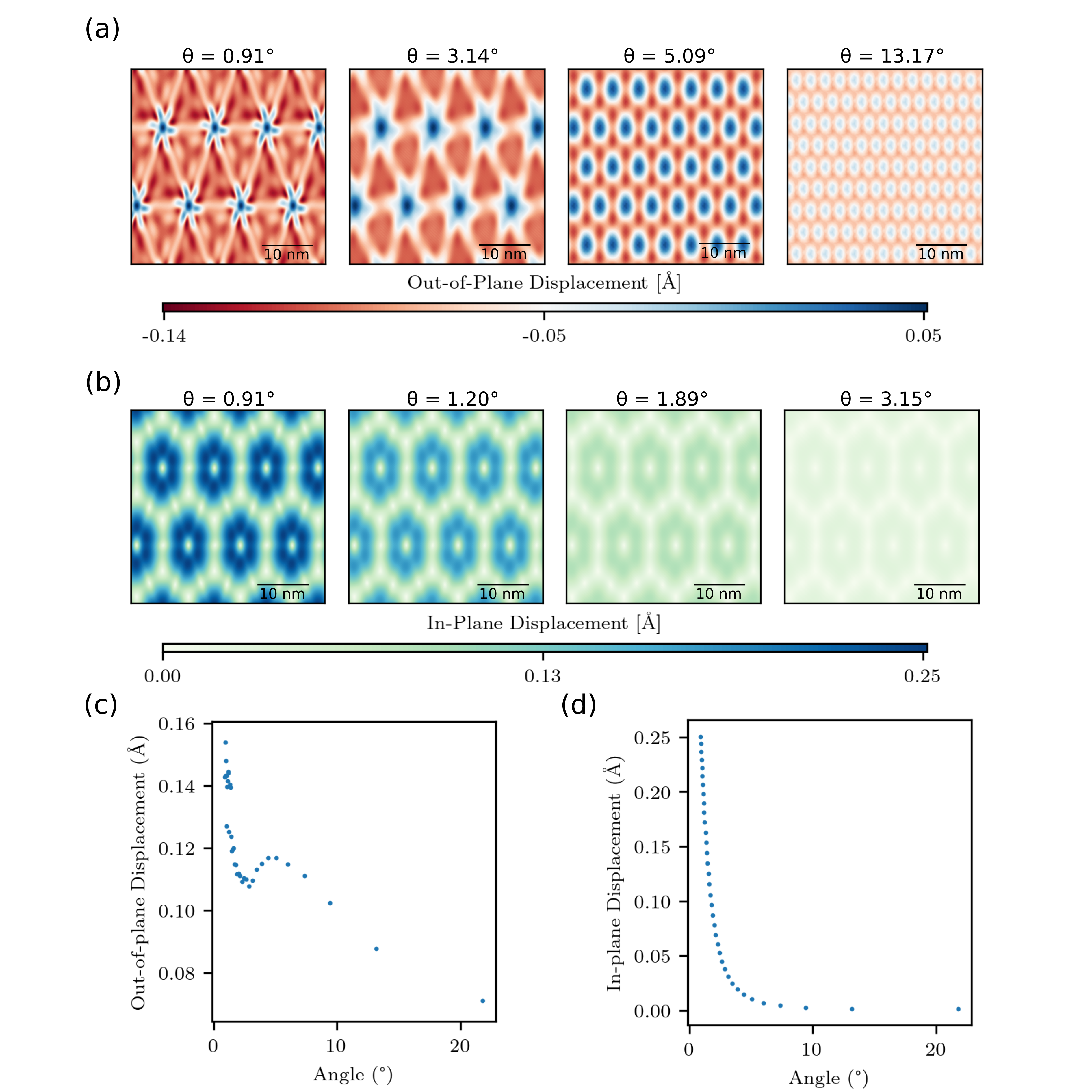}
    \caption{(a) Out-of-plane displacement map for four different twist angles after relaxation. (b) In-plane displacement map for four different twist angles after relaxation. (c) Maximum value of the out-of-plane displacement as a function of $\theta$. (d) Maximum value of the in-plane displacement, since some displacements are negative, the absolute value of this quantity is shown. In (a) and (b) only the view of the top layer is shown.}
    \label{fig:fig3}
\end{figure*}

Next, we calculate the structural properties of twisted $h$-BN, for which the results are presented in figures \ref{fig:interlayer_distance} and \ref{fig:fig3}. The primary descriptor for studying the structural properties is the interlayer distance, defined in this work as the average distance in the z-direction between each atom and its 3 nearest neighbors in the opposite layer. Since atoms in both layers do not necessarily overlap, an interpolation was applied to have a continuous description of the interlayer distance, shown in fig.\ref{fig:interlayer_distance}.b, from which we extract the moir\'{e} unit cell.
We can clearly observe four different regimes: moir\'{e} ($\theta$ > 13.17$^\circ$), transition (5.09$^\circ$ < $\theta$ < 13.17$^\circ$), soliton (1.89$^\circ$ < $\theta$ < 5.09$^\circ$), and domain-soliton($\theta$ < 1.89$^\circ$)  regimes\cite{arnold_relaxation_2023}. In the moir\'{e} regime we can model the twisted structures as rigidly twisted because the variation in interlayer distance becomes negligible. In the transition regime, relaxation effects become more important and some structural characteristics start to appear; the system starts going through a continuous transition into the soliton regime. In the soliton regime, solitons appear as bridges between nodes, and as borders for stacking domains. These stacking domains form a honeycomb lattice but are still small in comparison with the size of the nodes. When going further in decreasing $\theta$ we enter the domain-soliton regime, the stacking domains become dominant along the structure forming a hexagonal network with the nodes at the intersection points. In fig.\ref{fig:interlayer_distance}.c we can see the distribution of the interlayer distance as a function of the twist angle, with the dots representing the mean interlayer distance across the moir\'{e} unit cell, and the shaded region representing the interval between the maximum and the minimum interlayer distance. As we can see from the plot, the rigid twist approximation is valid at large twist angles, as the variation in interlayer distance becomes progressively smaller. We can see a steep decrease in $\langle d \rangle$, meaning the stacking domains become dominant, so we can associate this steep decrease as the shift from the transition regime to the soliton one. Similarly, in fig.\ref{fig:interlayer_distance}.d, we show the moir\'{e} periodicity $a_m$, which is obtained numerically from the interlayer distance landscape using our GAP model (blue dots) and compared to the analytical prediction (red line) \cite{latychevskaia_moire_2019}. We obtain an excellent match between both, further validating the accuracy of our model. 

We continue by analyzing the atomic displacements in structurally relaxed twisted layers of $h$-BN. The color map of the out-of-plane and in-plane displacements can be seen in fig.\ref{fig:fig3}.a and b respectively. The maximum value of each displacement as a function of $\theta$ are shown in fig.\ref{fig:fig3}.c-d. From these graphs, we can see that in the moir\'{e} regime there is no atomic reconstruction, since both in- and out-of-plane displacements are negligible. In the transition regime, the out-of-plane displacement gains relevance, but the in-plane displacement is still negligible. The in-plane displacement becomes significant with emerging solitons, while the out-of-plane displacement mirrors the interlayer distance landscape, peaking where the interlayer distance nodes appear. The small increase in the out-of-plane distance maximum at around 5$^{\circ}$, we attribute to a change of regime and is also in accordance with an increase in the interlayer distance interval, shown in fig.\ref{fig:interlayer_distance}.c.

Before discussing the polarization of twisted $h$-BN structures, it is essential to understand the origin of ferroelectricity and the resulting macroscopic polarization in these type of structures. The symmetry-breaking structural phase transition leads to the ordered arrangement of electric dipoles, resulting in spontaneous polarization that contributes to macroscopic polarization\cite{liu_van_2019}. Ferroelectricity spontaneously appears in twisted layers of $h$-BN from the overlap of B and N atoms, and the distortion of the $p_z$ orbitals of each atom. In previous work, it is shown that ferroelectricity is completely dictated by the commensurability of the system \cite{cao_interlayer_2022}. The AB/BA stacking configuration becomes dominant, hence maximizing the overlap of B and N atoms. We expect that in AB/BA stacking regions a spontaneous polarization emerges from the system. From the structural analysis of the system we see that AB/BA stacking regions become dominant after relaxation and at small $\theta$. There are several approaches to calculate the polarization of a system, ranging from the Berry Phase method\cite{spaldin_beginners_2012} to an approach in which we calculate the local density of states to obtain the charge distribution around the system, from which the electrostatic potential is obtained. We opt for the latter approach, since it is computationally cheaper while maintaining a high level of accuracy.

\begin{figure*}[!t]
    \centering
    \includegraphics[width=\linewidth]{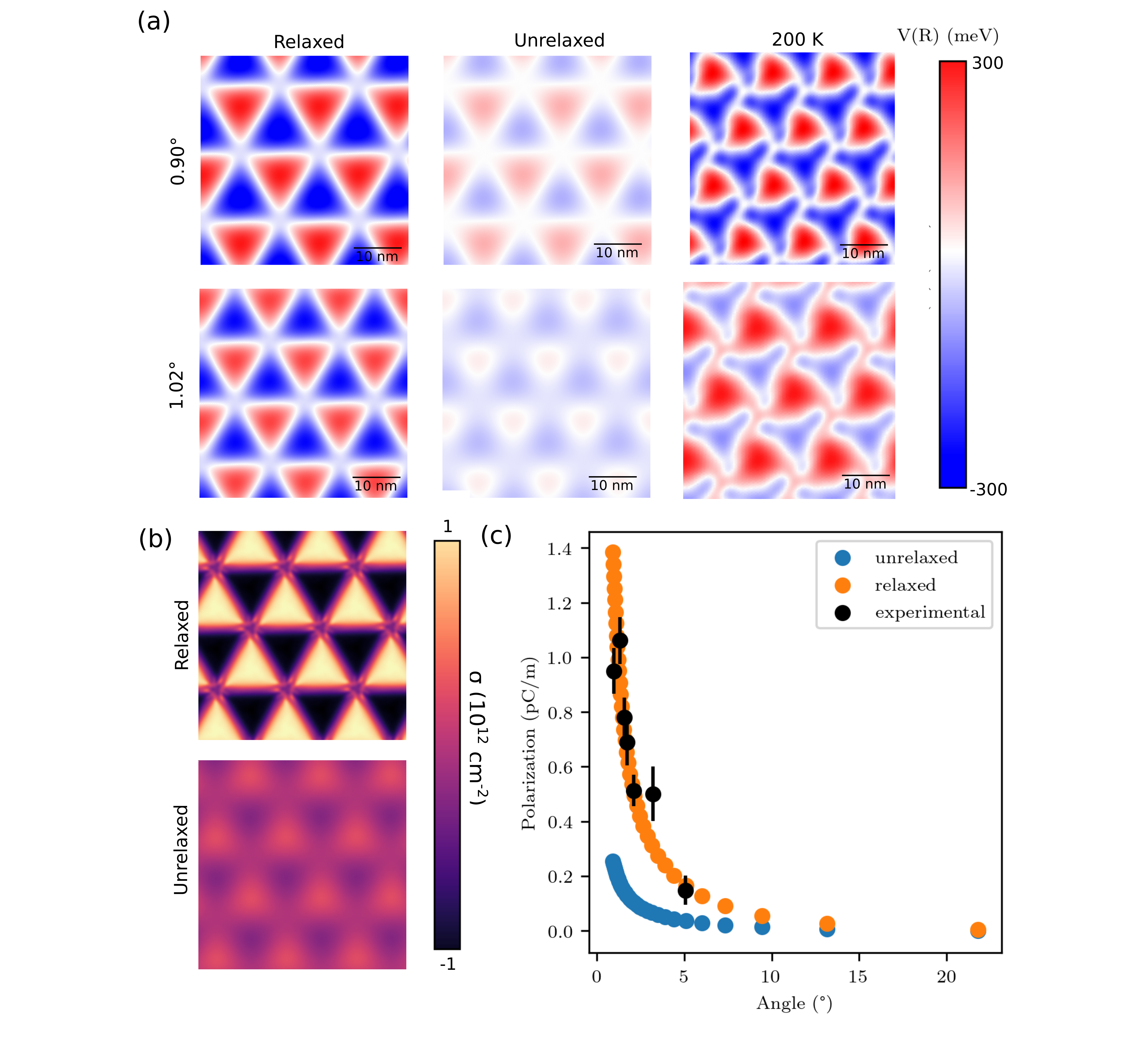}
    \caption{(a) Coulomb potential for two different twist angles, before and after relaxation, and after a MD run at 200 K. Only the top layer is shown. (b) Charge redistribution for a relaxed and a hard twist of the layers. The colorbar is in units of $10^{12} e cm^{-2}$. (c) $P_{max}$ as a function of $\theta$. The black dots represent experimental data with their corresponding errorbar obtained from \cite{kim_electrostatic_2024}. }
    \label{fig:fig4}
\end{figure*}

We set up a tight-binding model based on relative interatomic positioning that has proven valid in similar systems\cite{smeyers_strong_2023, moon_electronic_2014}. We used Pybinding \cite{moldovan_pybinding_2020}, a Python package, to set up and solve the tight-binding model, relying on the KPM method \cite{weise_kernel_2006} in order to obtain the electronic local density of states. For a more detailed description about the tight-binding model and the details about the calculation, we refer the reader to the Supplemental Information.

Experiments on twisted $h$-BN suggest that there is an electric potential pattern arising from the structure\cite{kim_electrostatic_2024}. We calculate the induced charge density distribution after twist by summing over all occupied states up to the charge neutrality point. As can be seen in fig.\ref{fig:fig4}.b, the relaxation emphasizes the triangular pattern that can be observed in experiments. Most of the induced charge is carried by B atoms, with the induced charge being $\sim$3 times larger than that of N atoms. In the soliton and domain-soliton regimes, the twist angle has the effect of increasing the size of the charged domains, but has no significant effect in the value of the electron density of the domains. The calculated electron density is on the order of $10^{12} cm^{-2}$, which is in accordance with previous theoretical results \cite{walet_flat_2021}. 

As mentioned above, twist induces an electrostatic potential at the surface of the $h$-BN top layer caused by the differential charge redistribution $\Delta \sigma$. This charge redistribution induces an electric polarization at the buried interface. Crystals of $h$-BN are naturally stacked in an AA' sequence, but in order to generate this polarization, the system prefers a more energetically favorable stacking configuration, AB/BA. The electrostatic potential is shown in fig.\ref{fig:fig4}.a, we can see that, similar to the charge redistribution, the electrostatic potential also has a triangular symmetry, and becomes neutral in regions of AA stacking as expected. The magnitude of the electrostatic potential also increases as the twist angle decreases, we attribute this to the increase in the size of the AB/BA domains, hence there is more twist-induced charge. 

The out-of-plane polarization of the system can be calculated following the procedure of \cite{zhao_universal_2021}. At close distance from the hBN Milorad Milosevicinterface we can define the electrical polarization as $P_{max} = \epsilon_0(V_{max} - V_{min})$, where $P_{max}$ is the electrical out-of-plane polarization at the AB stacking configuration, and the difference in the electrostatic potential is the potential drop when going from an AB stacking configuration to BA. In fig.\ref{fig:fig4}.c we can observe the value of the polarization as a function of $\theta$. Our results are in perfect agreement with previous experimental works\cite{kim_electrostatic_2024}, validating our methodology. As expected, the polarization is inversely proportional to $\theta$ because the size of the stacking domains follows the same trend which increases the amount of charge redistribution, thus giving a greater contribution to the polarization. We also see that atomic relaxation of the system enhances the out-of-plane polarization. In a previous theoretical work\cite{bennett_polar_2023} in which the out-of-plane polarization of bilayer hBN is calculated in the AB stacking, a value of 2 $[pC/m]$ is obtained. This is also the upper bound value we find for our calculations as we can see in fig.\ref{fig:fig4}.d. We also analyze the effect of temperature by repeating our calculation for systems that underwent a molecular dynamics run at temperatures 100 K, 200 K, 300 K, which are also included in our training set (see Supporting Information). The magnitude of the polarization remains unchanged at these temperatures, but as shown in fig.\ref{fig:fig4}.a the shape of the domains changes drastically. To analyze the MD at finite temperature, we ran the simulation for 30 ps and then did a statistical average with every step. We fixed the AA stacking nodes and proceeded with the average; we account for the change in the shape of domains to this pinning and the AB/BA regions moving around with the finite-temperature MD run.

To summarize, we have developed a machine learning potential using the Gaussian Kernel Regression method, achieving accuracy comparable to ab-initio calculations. We use our model to perform molecular dynamics simulations on moir\'{e} unit cells of twisted $h$-BN containing on the order of 10$^4$ atoms, a scale previously unattainable using ab initio methods, while also providing improved accuracy compared to standard classical potentials, even with applied uniaxial strain. The structural properties of the geometrically optimized structures obtained from our GAP model are consistent with the behavior of similar materials of the same class. Furthermore, we calculate the out-of-plane polarization of twisted $h$-BN using a tight-binding model based on interatomic distance-dependent hopping. Our results show excellent agreement with previous experimental results and improving upon the accuracy of previous ab-initio attempts. Since the polarization in twisted $h$-BN is of purely geometric origin, we can confirm that our model performs accurately at small twist angle. Additionally, our model captures the effects of finite temperature and agrees with experimental results showing minimal change in polarization with temperature. This approach opens a door to more effectively study the effects of electric fields in the nanoscale regime.

\section*{Acknowledgments}
This work was supported by the Research Foundation–Flanders (FWO–Vl), Special Research Funds of the University of Antwerp (BOF-UA), and the FWO–FNRS EOS project ShapeME. The computational resources and services for this work were provided by the VSC (Flemish Supercomputer Center), funded by the FWO and the Flemish Government – department EWI.

\bibliography{references.bib}

\appendix


\DeclareGraphicsExtensions{.png .jpg .pdf}

\makeatother

\author{Wilson Nieto Luna}
\affiliation{Department of Physics and NANOlight Center of Excellence, University of Antwerp, Groenenborgerlaan 171, 2020 Antwerp, Belgium}
\email{wilson.nieto@uantwerpen.be}
\author{Robin Smeyers}
\affiliation{Department of Physics and NANOlight Center of Excellence, University of Antwerp, Groenenborgerlaan 171, 2020 Antwerp, Belgium}
\email{robin.smeyers@uantwerpen.be}
\author{Lucian Covaci}
\affiliation{Department of Physics and NANOlight Center of Excellence, University of Antwerp, Groenenborgerlaan 171, 2020 Antwerp, Belgium}
\email{lucian.covaci@uantwerpen.be}
\author{Cem Sevik}
\affiliation{Department of Physics and NANOlight Center of Excellence, University of Antwerp, Groenenborgerlaan 171, 2020 Antwerp, Belgium}
\email{cem.sevik@uantwerpen.be}
\author{Milorad V. Milo\v{s}evi\'c}
\email{milorad.milosevic@uantwerpen.be}
\affiliation{Department of Physics and NANOlight Center of Excellence, University of Antwerp, Groenenborgerlaan 171, 2020 Antwerp, Belgium}

\title{Supporting Information for\\ {Machine-Learning Interatomic Potential for Twisted Hexagonal Boron Nitride: Accurate Structural Relaxation and Emergent Polarization}}

\date{\today}

\section{I. Generation of Training Set}

The accuracy of the potential depends on the robustness of the training set. In this purpose, we generated our training set based on ab-initio molecular dynamics(AIMD) using a Density Functional Theory (DFT) code as implemented in the Vienna Ab initio Simulation Package (VASP) \cite{kresse_ab_1994,kresse_efficient_1996}. In the following, all the ab-initio calculations were performed using the VASP plane-wave DFT code employing the Perdew-Burke-Ernzerhof (PBE)\cite{perdew_generalized_1996} functional with the DFT-D3 dispersion correction method \cite{grimme_consistent_2010} using the Becke-Johnson damping \cite{grimme_effect_2011}, an energy cut-off of 620 eV, a Gaussian smearing of 0.05 eV, and projector augmented wave pseudopotentials \cite{blochl_projector_1994,kresse_ultrasoft_1999} were also employed. We chose the PBE+D3 correction because it was successfully used in a previous GAP potential for hBN in several phases \cite{thiemann_machine_2020}, and also it has proven successful in predicting lattice parameter of bulk hBN. The reciprocal lattice lattice was sampled in the Monkhorst-Pack grid \cite{monkhorst_special_1976} with a maximum distance between the k-points of 0.02 \r{A}$^{-1}$. 

\begin{figure*}[t]
    \centering
    \includegraphics[width=\linewidth]{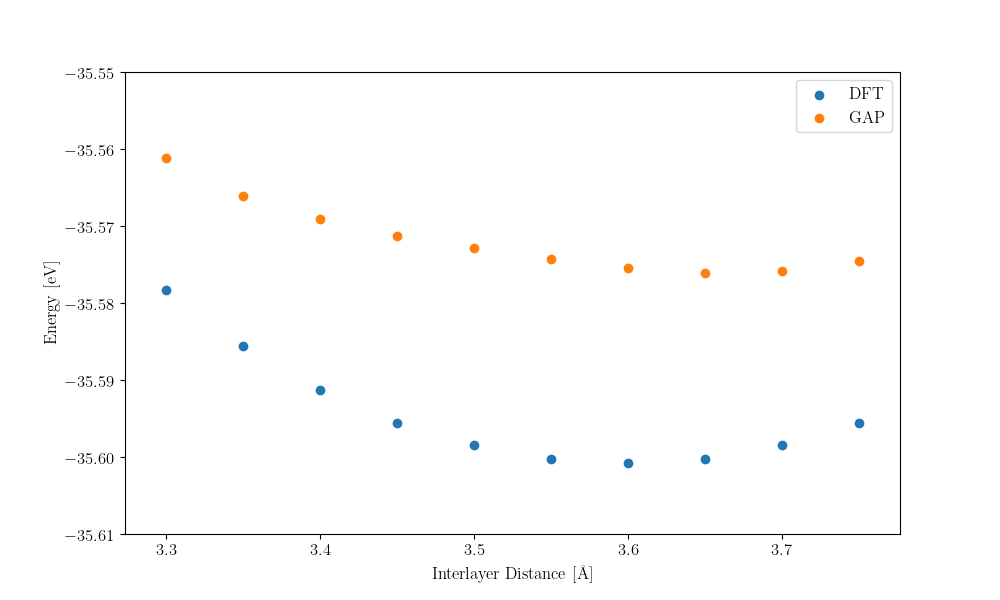}
    \caption{Equation of state for bilayer $h$-BN in AA stacking including the comparison with the prediction of our GAP potential.}
    \label{fig:s1}
\end{figure*}

Before generating the twisted structures to perform AIMD on them, we have to find the appropriate interlayer distance according to the settings of our simulation, this is shown in figure \ref{fig:s1}. For the lattice constant we take the established value for hBN from previous works\cite{lynch_effect_1966}. After establishing these values, we proceeded and generated twisted structures of hBN using the Twister python package \cite{naik_twister_2022,naik_ultraflatbands_2018}, creating bilayer hBN twisted structures from 0.91$^\circ$ to 21.78$^\circ$, all according to the definition of commensurate unit cells\cite{pal_twist_2024}. 

To construct the data base, we followed a similar procedure as the one described in \cite{rowe_development_2018}, in which we build the training set using an iterative approach. We started with an AIMD run at 100 K for angles of 21.78$^\circ$, 13.17$^\circ$, 9.43$^\circ$ and 7.34$^\circ$, running these for at least 1000 fs to obtain enough samples of the configuration space. From all these configurations,  we choose ~1000 configurations to train the first version of our GAP based on the energies and forces for each atom. With this first version of our potential we generate different MD trajectories at 100 K, 200 K and 300 K, and do a single point energy calculation to obtain energies and forces from each individual atom. We proceed to train once again a different version of the GAP and repeat this process iteratively until we achieve the desired accuracy. The last training set contained 1300 different configurations of MD trajectories and relaxation trajectories, and from this we chose 20\% of them for a test set.

\newpage

{\section{II. Gaussian Approximation Potential}

Gaussian Approximation Potential (GAP) was proposed as a solution to the problem of interpolating the Born-Oppenheimer potential energy surface (PES) of a system via the application of the Gaussian kernel regression machine learning methodology \cite{bartok_gaussian_2010}. The advantage of GAP over previous empirical potentials is that it does not assume a functional form into which the PES has to be decomposed. From the ab-initio dataset, GAP has only available the energies, forces and virial stresses of the system, so in that sense, to facilitate systems with larger sizes, GAP decomposes the atomic environment into a sum of local contributions. These contributions are then computed via kernel functions which represent the similarity between chemical environments. Usually we decompose the contributions into two body, three body, or many body interactions, specifically for this work we decided to represent every contribution with many body interactions. 

The many body interaction takes the form of the smooth overlap of atomic positions (SOAP) descriptor, within this descriptor, each atomic environment is described by a local density of
neighbors, generated by summing over Gaussians that are placed in each atom within a specific cut-off. The density for the $i$th atom is expanded in a basis of radial functions and spherical harmonics: 

\begin{align}
    \rho_i(\boldsymbol{r}) = \sum_{\substack{n < n_{max} \\ l < l_{max} \\ |m| \leq l }} c^{i}_{nlm} g_{n}(r) Y_{lm} (\boldsymbol{r}) \label{eq:soap} 
\end{align}

wherein the power spectrum of the coefficients $c^{i}_{nlm}$ form the SOAP vectors. It is worth mentioning that these SOAP descriptors are invariant under rotations, which becomes relevant for our system of study. 

In addition to the descriptor, there are several other hyperparameters that impact the performance and accuracy of a GAP potential; in this work we will only focus on the most important ones, for a more detailed discussion about GAP descriptors and hyperparameters, consult elsewhere \cite{klawohn_gaussian_2023}.  Due to the flexibility of its descriptors
and hyperparameters, GAP is the most promising ML potential to reproduce the vdW interaction in layered heterostructures of 2D materials. GAP also has the clear advantage of low computational cost with respect to ab-initio methods, which is critical when dealing with a large number of atoms (such as in so-called moiré heterostructures, where periodic unit cells become very large due to lattice mismatch or twist between the constituent monolayers). In table \ref{table:model parameters} the relevant hyperparameters are shown, we chose a bigger cut-off than the one from the previous GAP potential for hBN in order to also include the interlayer interaction in this descriptor, and in this way discard the use of another descriptor to reproduce different interactions. Due to the trade-off between computational cost and accuracy of the model, we decided to truncate the spherical harmonics of the SOAP descriptor after the 6th order, this has proven to be enough to reach the desired accuracy. Instead of using all of the local environments of the training set, a covariance matrix is constructed based on 1100 points, this will focus the model to train on specific parts of the training configuration space. The target deviations assign the fit and the weight between the values predicted from the model to the target values in the training configuration. 

\begin{table}[t]
\centering
\begin{tabular}{|c|c|}
\multicolumn{2}{c}{\textbf{SOAP descriptor}} \\
\hline
cut-off [\r{A}] & 6.0 \\
cut-off width [\r{A}] & 1.0 \\
$\delta$[eV] & 0.2 \\
sparse method & CUR \\
sparse points & 1100 \\
$l_{max}$ & 6 \\
$n_{max}$ & 6 \\
$\zeta$ & 2 \\
\hline
\multicolumn{2}{c}{\textbf{target deviations}} \\
\hline
$\sigma_{energy}$ &  0.005\\
$\sigma_{force}$ &  0.0005\\
$\sigma_{virial}$ &  0.005\\
\hline
\end{tabular}
\caption{Model parameters for the twisted bilayer hBN GAP.}
\label{table:model parameters}
\end{table}
%


\newpage
\section{III. Tight Binding Model}

We describe the system with a tight-binding model that uses Bloch wave functions of the form

\begin{align}
    |\Psi_{\textbf{k}}\rangle = \frac{1}{\sqrt{N}} \sum^{N}_i e^{i\textbf{k}\textbf{R}_i} |\phi_i \rangle \label{eq:bloch function} 
\end{align}

where $|\phi_i \rangle$ describes the wavefunction pf atomic site $i$ at position $\textbf{R}_i$. The general form of the tight-binding Hamiltonian is

\begin{align}
    H = -\sum_{i,j} t(\textbf{R}_i - \textbf{R}_j) |\phi_i \rangle \langle |\phi_i| + \sum_i V(\textbf{R}_i) |\phi_i \rangle \langle |\phi_i|  \label{eq:hamiltonian} 
\end{align}

In the Hamiltonian from eq. \ref{eq:hamiltonian}, the first term defines the hopping between sites $i$ and $j$ with strength $t$. The second term represents the on-site potential, in which we used $V_B = -1.287$ eV and $V_N = -5.393$ eV, values obtained from \cite{javvaji_ab_2025}. We used the Slater-Koster type of functions \cite{slater_simplified_1954}, which give a good estimate of the hopping value depending on the relative positions between atomic orbitals. Following the procedures from \cite{moon_electronic_2014} we get to the following expression for the distance depending hopping:

\begin{align}
    t(r_{ij}) = \left( \frac{z_{ij}}{r_{ij}}\right)^2 V _{pp\sigma}(r_{ij}) + \left(1 - \left(\frac{z_{ij}}{r_{ij}}\right)^2\right)V_{pp\pi}(r_{ij})  \label{eq:hopping} 
\end{align}
 where $z_{ij}$ is the component of $r_{ij}$ along the z-axis. We also define

\begin{subequations}

\begin{align}
     V_{pp\sigma}(r_{ij}) = \gamma_1 \text{ exp}\left({q_\sigma \left(1 - \frac{r_{ij}}{c}\right)}\right) \\
     V_{pp\pi}(r_{ij}) = \gamma_0 \text{ exp} \left( {q_\pi \left(1 - \frac{r_{ij}}{a_{BN}}\right)} \right) \\ 
     \frac{q_\sigma}{c} = \frac{q_\pi}{a_{BN}} = \frac{ln(\gamma_0^{'} / \gamma_0)}{a_{BN} - a}\label{eq:potentials}
\end{align}

\end{subequations}

where $\gamma_0$ is the nearest neighbor interaction (-2.7 eV), $\gamma_0^{'}$ is the next nearest neighbor interaction (0.1$\gamma_0$), $c$ parameter is the interlayer distance in the AA stacking, and $a$ is the optimized lattice parameter. We have to define specific hopping parameter prefactors depending on the atomic species, so we defined them as

\begin{subequations}

\begin{align}
     \gamma_{BB,1} = 0.831 \text{eV}\\
     \gamma_{NN,1} = 0.3989 \text{eV} \\ 
     \gamma_{BN,1} = 0.6601 \text{eV}
     \label{eq:prefactors}
\end{align}

\end{subequations}


%
%
%
%
%
%
%
%
%
%
%
%

\end{document}